\documentclass[aps,prx,showkeys,twocolumn,superscriptaddress]{revtex4}
\usepackage{amssymb,amsmath}
\usepackage{bm,graphicx}
\usepackage{array}
\graphicspath{{Pictures/}}
\usepackage{natbib}
\usepackage{subfigure}
\usepackage{mathptmx}
\usepackage{pxfonts}
\usepackage{times}
\usepackage[T1]{fontenc}
\usepackage{epstopdf}
\usepackage[colorlinks,citecolor=blue,linkcolor=cyan]{hyperref}
\usepackage[usenames,dvipsnames,svgnames]{xcolor}

\begin{document}

\title {Anomalous Dynamical Responses in a Driven System}

\author{Suman Dutta}
\email{Email:sumand@bose.res.in}
\affiliation{Department of Chemical Biological and Macromolecular Sciences\\
S. N. Bose National Centre for Basic Sciences \\ Block-JD, Sector-III, Salt Lake, Kolkata 700098 \\ India.\\}

\author{J. Chakrabarti}
\email{Email:jaydeb@bose.res.in}
\affiliation{Department of Chemical Biological and Macromolecular Sciences\\
S. N. Bose National Centre for Basic Sciences \\ Block-JD, Sector-III, Salt Lake, Kolkata 700098 \\ India.\\}

\affiliation{The Thematic Unit of Excellence on Computational Materials Science,
S. N. Bose National Centre for Basic Sciences}

\date{15 July, 2016 }

\begin{abstract}
The interplay between structure and dynamics  in non-equilibrium steady-state is far from understood. We address this interplay by tracking Brownian Dynamics trajectories of particles in a binary colloid of opposite charges in an external electric field, undergoing cross-over from homogeneous to lane state, a prototype of heterogeneous structure formation in non-equilibrium systems. We show that the length scale of structural correlations controls heterogeneity in diffusion and consequent anomalous dynamic responses, like the exponential tail in probability distributions of particle displacements and stretched exponential structural relaxation. We generalise our observations using equations for steady state density which may aid to understand microscopic basis of heterogeneous diffusion in condensed matter systems.\\
\end{abstract}


\maketitle

 The dynamic response depicts how disturbance in any thermodynamic quantity in a system relaxes with time via particle motion \cite{hm,cl}. This forms the microscopic basis of transport processes\cite{hm,cl,soft1,soft2}, like diffusion in liquids. In general particle dynamics depends on the structure of the system.   Non-equilibrium systems often show emergence of structural heterogeneity in steady states, known as pattern formation\cite{hoh} relevant in areas ranging from material science to biological systems \cite{soft1,soft2}. The particle dynamics in such systems is nontrivial due to drive \cite{nat3,nat3a,nat4,nat4a}.  Despite large number of studies \cite{soft1,soft2,hoh,nat3,nat3a,nat4,nat4a}, the connections between dynamics and structure, and consequently, the transport processes are not well understood for non-equilibrium situations. This motivates us to explore dynamic responses and relate them to the underlying structural morphology in non-equilibrium steady states.     
       
 Colloids are ideal model system to realize condensed matter properties both in and out of equilibrium\cite{soft1,soft2,nat3,nat3a}. Variety of structures can be induced in colloids by external perturbations\cite{soft1,soft2,nat3,nat3a,nat4,nat4a}. Moreover, the particles are big enough for optical imaging and are slow so that the particle motions can be followed\cite{hm,cl,soft1,soft2,nat3,nat3a,nat4,nat4a}.  Several experiments and theoretical works  show that in an external uniform electric field of large strength drives a binary mixture of oppositely charged colloids to form heterogeneous structure in the steady state with lanes of dynamically locked-in like-charged particles, while the mixture is homogeneous at low field \cite{soft1,soft2,nat3,nat3a,nat4,nat4a,nat5,nat5a,nat6,nat6a,nat6b,nat7a,nat7b}. 
      
      We consider steady states of binary charged colloid in electric field. The steady state structural correlations are given by the pair correlation functions(PCF) \cite{hm} which are probability distributions of particle separations at a given time. We study how density changes relax via microscopic motion in different steady state structures of binary charged colloid in electric field. The density relaxation, also known as the van Hove function (vHf) \cite{vhv}, is measurable from scattering experiments and routinely used in theoretical studies to probe dynamic responses \cite{hm,cl,vhv}. The vHf consists of two parts: (1)The probability distribution of displacements of individual particles in a given time interval (self-vHf), characterising particle motions; and (2) that of the separations between pair of particles (distinct-vHf), describing relaxation of structural changes. The self-vHf for a normal liquid is Gaussian whose width increases linearly with time, the proportionality constant defining the self-diffusion co-efficient, D\cite{vhv}. The distinct-vHf decays exponentially in time with a rate $Dq^{2}$ where $2\pi/q$ is characteristic length of density changes\cite{hm}. Both PCF and vHfs can be calculated from particle coordinates.  We compute particle coordinates using the Brownian Dynamics trajectories of driven colloidal particles. The PCFs show that the structural correlations grow with increasing electric field. The system has normal liquid-like dynamic responses both in homogeneous and lane phases. At intermediate field the self-vHf develops a slow exponentially falling spatial tail and temporally a stretched exponential structural relaxation. The single particle motions show that the anomaly is due to heterogeneity in self-diffusion of particles, having non-monotonic dependence on structural correlations. We come up with a model to explain this non-monotonicity. 
 
 We show the presence of slow dynamics even though the system is subject to strong drive. This slowing down of dynamics is different from that in super-cooled liquids\cite{smk,slow,cage}. While the particles in super-cooled systems show caging by the neighboring particles, the particle dynamics in our system is diffusive. The slow dynamics, here, is associated with heterogeneity in diffusion. Moreover, the heterogeneity can be tuned by drive. This points out to novel transport mechnaism in a driven system. More importantly, the generalization presented via the steady state model may be useful to understand microscopic origin of heterogeneous diffusion in a variety of condensed mater systems\cite{ng} . 
 
Our system consists of a binary mixture of equal number of positively ($N_{+}$) and negatively ($N_{-}$) charged colloidal particles of diameter $\sigma (=1\mu m$)($N_{+}$=$N_{-}$=1000) in a solvent fluid of viscosity $\eta (=1cP)$ in a cubic box of length ($L=21.599\sigma $) at temperature $(T=298K)$ with the periodic boundary conditions. The pair interactions between the particles \cite{nat6a} at positions $\mathop{R_{i}}\limits^{\to} $ and $\mathop{R_{j}}\limits^{\to } $ with separation $r_{ij} =|\mathop{R_{i}}\limits^{\to } -\mathop{R_{j}}\limits^{\to } |$, $V(r_{ij} )= V_{SC} (r_{ij})+ V_{Repulsion} (r_{ij})$ with
$V_{SC} (r_{ij}) = V_{0} [q_{i} q_{j}/(1+\frac{\kappa \sigma }{2})^{2}] [\exp(-\kappa \sigma ((r_{ij} /\sigma )-1))/(r_{ij} /\sigma)]$
and $V_{Repulsion} =\varepsilon [(\sigma /r_{ij} )^{12} -(\sigma /r_{ij} )^{6} ]+\frac{1}{4}$ for $r_{ij} <2^{1/6} \sigma$ and zero, elsewhere. Here  $q_{i}$ the charge of the $i$th particle, $\kappa $ the inverse screening length, $V_{0 }$ the interaction strength parameter and $\varepsilon =4\left|q\right|^{2} V_{0} (1+\kappa \sigma /2)^{2} $ with $q_{i}=q_{j}=q$ \cite{nat6a}. We fix $\kappa \sigma (=5.0)$and $V_{0}^{*} =\left|q\right|^{2} V_{0} /k_{B} T(=50.0)$ as in Ref. \cite{nat6a}.

 The BD simulations are carried out using discretized form of the Langevin's equation \cite{erm} with large viscous damping $\Gamma(=3\pi\eta\sigma)$, electric field $f_{0}$ in z-direction and fluctuating force $\vec{F}$ having variance $<F^{\alpha } (t)F^{\beta } (t')>=2D_{0} \delta _{\alpha \beta } \delta (t-t')$. Here $\alpha $,$\beta $ denote the cartesian components and $D_{0}(=k_{B}T/3\pi \eta \sigma$, $k_{B}$ the Boltzmann constant) the Einstein-Stokes Diffusion coefficient. We take $\tau _{\beta }( =\sigma ^{2} /D_{0} $ ) as unit time, $\sigma$ the length unit and $k_{B} T$ the energy unit. The integration time step $\Delta t=0.00005$. The analysis is carried out in the steady state for a non-zero $f(=\left|q\right|f_{0} \sigma /k_{B} T$), applied after the system is equilibrated without electric field.
 
 We characterize structures by single particle densities \cite{hm}, for both species in XY and XZ planes $\rho ^{(\pm)} (X,Y)$ and $\rho ^{(\pm)} (X,Z)$ respectively. The steady state is also characterized via a lane order parameter, $\Phi$ defined in Ref. [21]. The self-vHf for ions displaced by a distance $\vec{\Delta r} (=\Delta z,\vec{\Delta r_{\bot}} $) in time interval $t$ [1], $G_{S}^{(\pm )} (\mathop{\Delta r}\limits^{\to } ,t)=(1/N_{\pm } )<\sum _{i=1}^{N_{\pm}}\delta (\mathop{\Delta r}\limits^{\to } +\mathop{R_{i}}\limits^{\to } (t)-\mathop{R_{i}}\limits^{\to } (0) >$. The distinct-vHf for the pair of $+ve$ ions separated at a distance $\vec{r} (=z,\vec{r_{\bot}}$ in time interval $t$ is given by  $G_{D}^{(++ )} (\vec{r} ,t)=(2/N_{+}(N_{+}-1))<\sum _{i,j=1}^{N_{+}}\sum _{j\neq i}\delta (\vec{r} +\vec{R_{j}}(t)-\vec{R_{i}} (0) >$ while that between a pair of $+ve$ and $-ve$ charges is $G_{D}^{(+-)} (\vec{r} ,t)=(1/N_{+}N_{-} )<\sum _{i=1}^{N_{+}}\sum _{j=1}^{N_{-}}\delta (\vec{r} +\vec{R_{j}}(t)-\vec{R_{i}} (0) >$. Similarly, the distinct-vHf for the $-ve$ particles are defined. All the vHfs are computed by averaging over initial conditions and different Brownian trajectories starting from different equilibrium configurations. The PCFs for the +ve ions $g_{f} ^{(++)} (\mathop{r_{\bot}}\limits^{} ,z)_{} $, the --ve ions $g_{f} ^{(--)} (\mathop{r_{\bot}}\limits^{} ,z)_{} $and between +ve and --ve ions $g_{f} ^{(+-)} (\mathop{r_{\bot}}\limits^{} ,z)_{} $, are given by the $t=0$ value of the respective distinct-vHfs.  We compute the in-plane probability distribution of the particle displacements in time of 40 randomly tagged particles, $P_{}^{(+)(i)} (\mathop{r_{\bot}}\limits ,t)$ sampled over 30 Brownian trajectories.

Both species behave similarly, and we focus on the +ve species. Nearly homogeneous mixed phase, shown by  density profiles $\rho^{(+)} (X,Y)$ in XY plane and $\rho^{(+)} (X,Z)$ [Fig.1] in XZ plane for \textit{f}$=50$ with tiny domains of like charged particles elongated parallel to the field [Figs.1(a-b)], evolves through a pre-lane state having much bigger domains with increasing \textit{f}$ (= 150)$ [Figs. 1(c-d)]. Finally the lane state takes place, as in earlier observations \cite{nat7a}, for sufficiently large \textit{f }$(=300)$, having network of large domains in XY plane along with vertical lanes in the XZ plane [Figs. 1(e-f)]. Some snapshots of particle configurations for different $f$ are given in SI Fig.S1. The order parameter $\Delta \Phi[=\Phi(f)-\Phi(0)]$ \cite{nat6a} for laning shows continuous rise with $f$ [SI Fig. S2]. The structural morphologies in XY plane are very similar to those in pattern forming liquids \cite{hoh}. 

\begin{figure}[h]
\includegraphics[angle=0,scale=0.6]{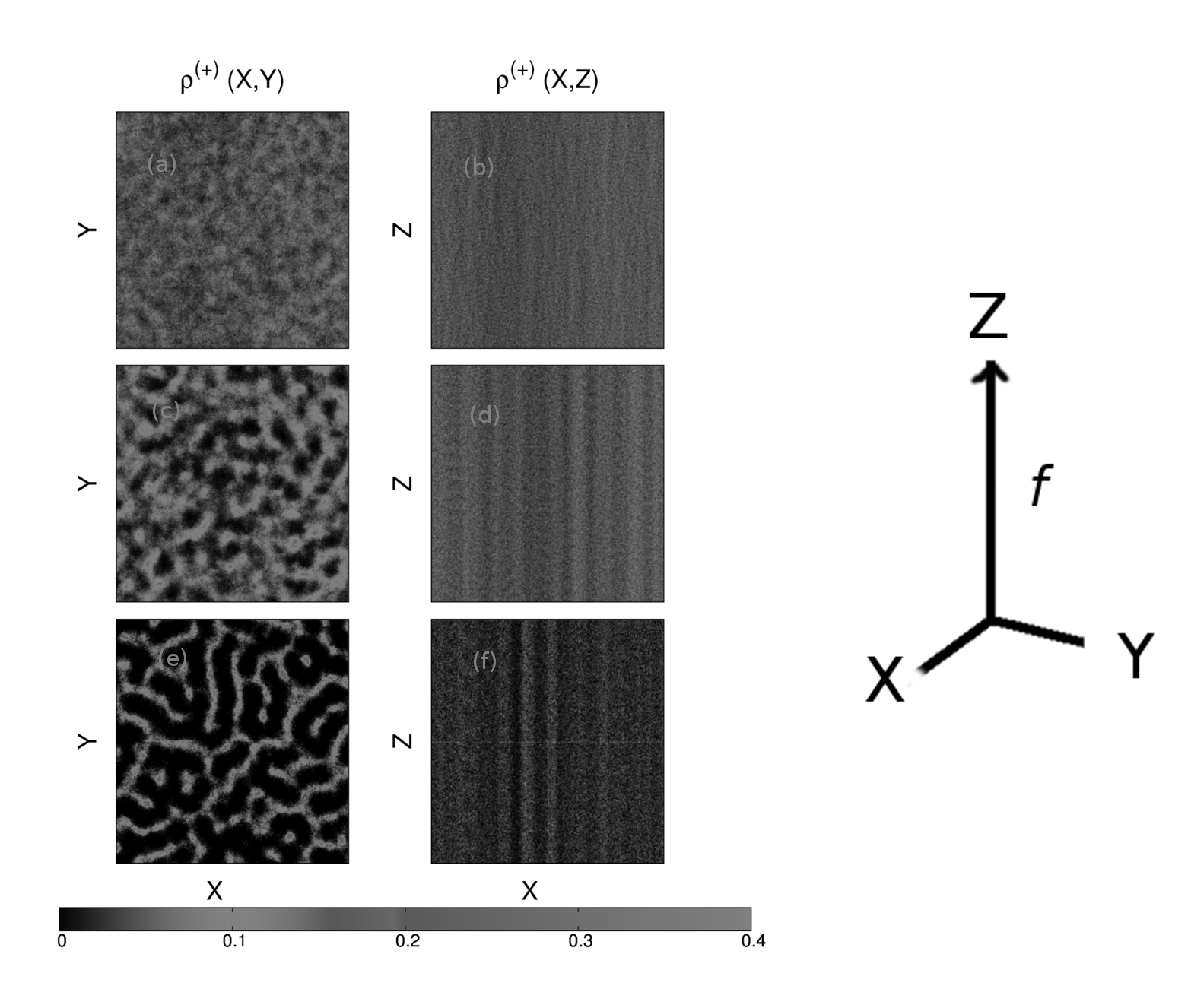}
\caption{Density Plots (a) $\rho ^{+} (X,Y)$ and (b) $\rho ^{+} (X,Z)$for \textit{f}=50; (c) $\rho ^{+} (X,Y)$ and (d) $\rho ^{+} (X,Z)$for \textit{f}=150; and (e) $\rho ^{+} (X,Y)$ and (f) $\rho ^{+} (X,Z)$for \textit{f}=300. The relevant direction including the field ($f$) direction, $Z$ are indicted in the figure. The contour values are indicated by the shades at the bottom}
\end{figure}

{\it Structural Correlations:} The PCFs, $g_{f} ^{(++)} (\mathop{r_{\bot}}\limits ,z)$  between $+ve$ species and $g_{f} ^{(+-)} (\mathop{r_{\bot}}\limits ,z)$ between $+ve$ and $-ve$ species in terms of particle separation $r_{\bot}$ in the plane transverse to the field and that parallel to field, $z$ are shown in Fig.2. We observe correlations  only upto single particle diameter for \textit{f}$=50$[Inset, Fig.2(a)]. At $f=150$, the correlations in $g_{f} ^{(++)} (\mathop{r_{\bot}}\limits ,z)$ extend up to a couple of coordination shells. The strong peak in $g_{f} ^{(+-)} (\mathop{r_{\bot}}\limits ,z)$ for $\mathop{r_{\bot}}\limits \approx 1$, indicate tendency of alignment of the positively charged particles in vertical lanes with short ranged correlations in the transverse plane in the pre-lane state (Fig.2(a)). Fig.2(b) reveals PCFs for fully developed lane phase with enhancement both in vertical and in-plane correlations extending upto several particle diameter at higher \textit{f $(=300)$.} Thus length scale of structural correlations increases with \textit{f.} 

The correlation energy, given by \cite{hm} $E^{C+} (f)=\int V^{(++)} (r)g_{f} ^{(++)}  (r_{\bot } ,z) d^{2} r_{\bot } dz$ + $\int V^{(+-)}$ $(r)g_{f} ^{(+-)}$ $(r_{\bot } ,z) d^{2} r_{\bot } dz$, is the cost of internal energy for bringing the positively charged species in a domain, replacing the negatively charged species. Fig. Inset, 2(b) shows the correlation energy $\Delta E^{C+} =E^{C+} (f)-E^{C+} (0)$ and energy due to the external electric field, $E^{D+} (f)=2fq\int _{0}^{L/2}z\rho (z) dz$ as functions of \textit{f}.  The energy cost of bringing the same charges in a domain is compensated by the external electrostatic energy above $f_{C}= 200 $, leading to fully developed lanes of like charged particles. 
\begin{figure}[h]
\includegraphics[angle=0,scale=0.07]{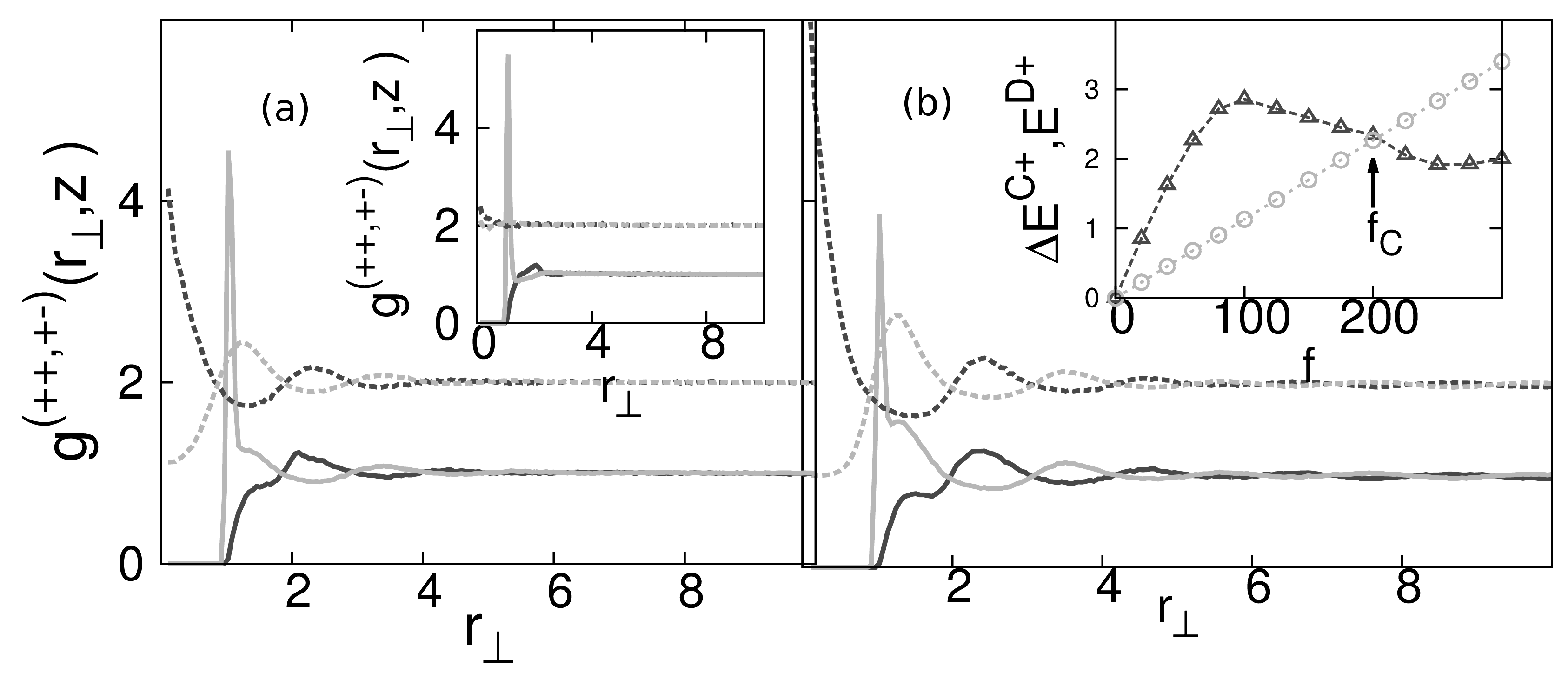}
\caption{Structural Correlations: $g_{f} ^{(++)} (\mathop{r_{\bot}} ,z)_{} $ (black) and $g_{f} ^{(+-)} (\mathop{r_{\bot}} ,z)_{} $ (grey) as function of $\mathop{r_{\bot}} $ for z=0 and z=10.7 (with vertical offset 0.5) for (a) $f=150$ and Inset, $f=50$ and (b)$f=300$. Inset, $\Delta E^{C+} (f)$ (triangles) and $E^{D+} (f)$ (circles) as functions of $f$.}
\end{figure}

{\it Dynamic Responses:} The self-vHf for displacement in $\pm z$ direction, $\Delta z$ and the transverse plane, $\Delta r_{\bot}$ are different, $G_{S}^{(\pm )} (\Delta z,t)\ne G_{S}^{(\pm )} (-\Delta z,t)\ne G_{S}^{\left(\pm \right)} \left(\Delta r_{\bot } ,t\right)$ for $f \neq 0$. We focus on in-plane quantities where the structural morphologies show distinct changes ( $\Delta z$ data given in SI Figs.S3 and S4). For small \textit{f}(=50) $G_{S}^{(+)} (\mathop{\Delta r_{\bot}}\limits ,t)$ are  Gaussian, shown in Fig.3 (a). Such Gaussian self-vHf is a characteristic of a normal liquid\cite{hm}.  However, $G_{S}^{(+)} (\mathop{\Delta r_{\bot}}\limits ,t)$ develop spatially exponential decay tails for large $\Delta r_{\bot}$  [Fig. 3(b)]. The amplitude of the Gaussian part relative to that of the exponential tail for large $t$ approaches the ratio $\Phi(f) /(1-\Phi(f))$. This implies that exponential tail develops due to $+ve$ particles in the neighborhood of $-ve$ particles. The self-vHfs regain Gaussian form but with double peaks for \textit{f }(=300)$>f_{C}$ (Inset Fig.2(b)) in the fully developed lane phase [Fig. 3(c)].

\begin{figure}[h]
\includegraphics[angle=0,scale=0.15]{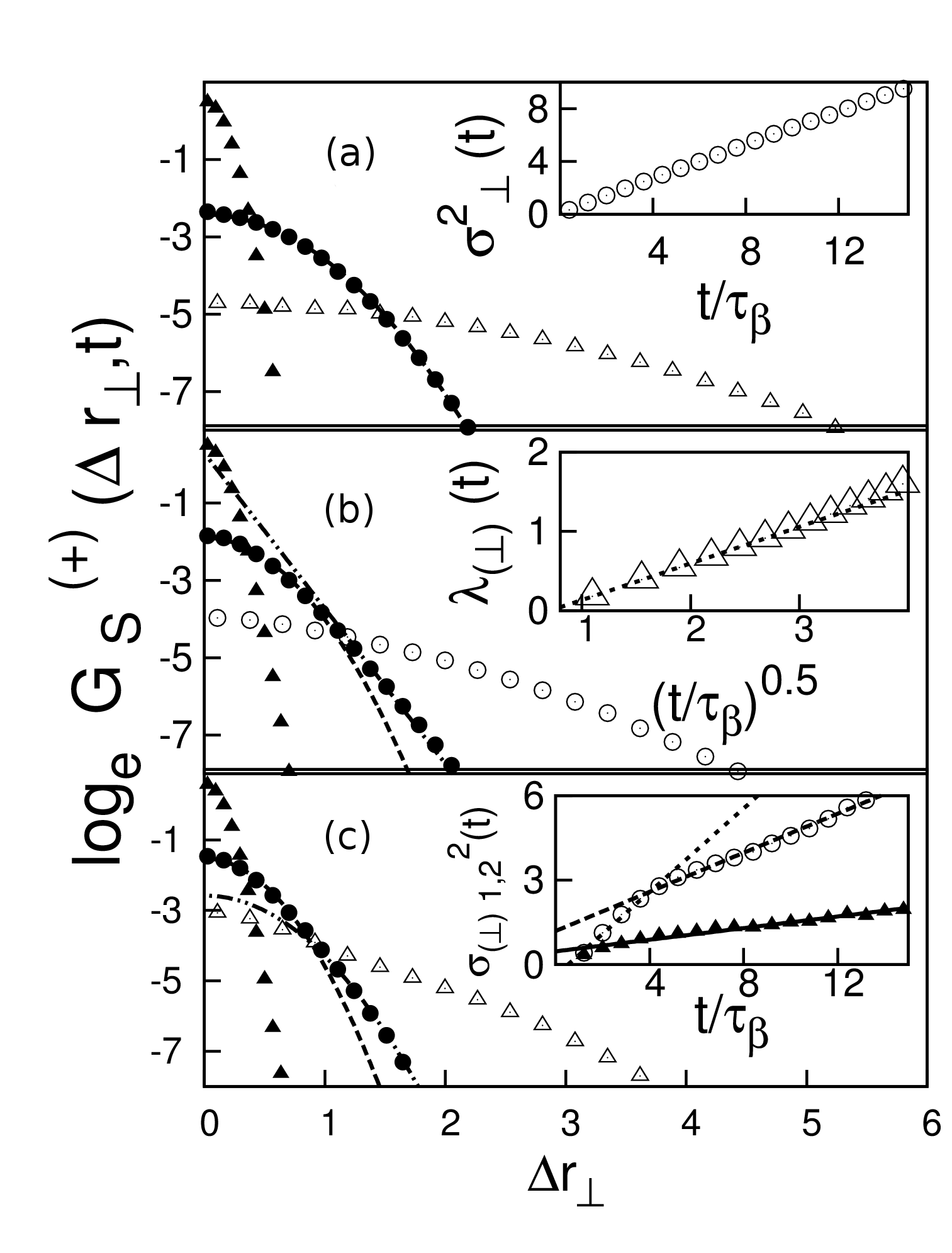}
\caption{Dynamical quantitites: 
(a)   ln$G_{S}^{(+)} (\mathop{\Delta r_{\bot}} ,t)$ vs $\Delta r_{\bot } $ for \textit{f}$=50$; Dashed line: Gaussian fit. Inset: $\sigma^{2}_{\bot} (t) $ as a function of $t$ (b) ln$G_{S}^{(+)} (\mathop{\Delta r_{\bot}},t)$ as a function of $\Delta r_{\bot } $ for \textit{f}$=150$;  Dashed line: Gaussian and  Dot-dashed line: exponential tail. Inset: $\lambda_{\bot} (t)$ as a function of $t$ (triangles), Dotted line shows $\lambda_{\bot} (t)$ $\sim $ $(\frac{t}{\tau_{\beta}})^{0.5}$.(c)$ln G_{S}^{(+)} (\mathop{\Delta r_{\bot}} ,t)$for \textit{f}$=300$ ; Dashed line and Dot-dashed line: Double Gaussian fit. Inset: Dependences of $\sigma_{\bot (1)}^{2} (t) $ and $\sigma_{\bot (2)}^{2} (t) $  on $t$.}
\end{figure}

The changes in the self-vHf take place at critical values, $\Delta r_{\bot}=r_{c}$.  We fit $G_{S}^{(+)} (\mathop{r_{\bot}}\limits^{},t)$ $\sim$ $\exp (-r_{\bot } ^{2} /\sigma _{\bot } ^{2} (t))$ for $r_{\bot } <r_{c} $ and $\exp (-r_{\bot } /\lambda _{\bot } (t))$ for $r_{\bot } >r_{c} $ for $f=150$. We also fit the data for $f=300$ with Gaussians with width parameters $\sigma _{\bot (1) }^{2} (t)$ for $r_{\bot } <r_{c} $ and $\sigma _{\bot  (2)}^{2} (t)$ for $r_{\bot } >r_{c} $. For $f=150$ and $f=300$, we show the dependence of $r_{c}$ as a function $t$ in Fig. SI Fig. S5 and S6 respectively. We find that \textit{r${}_{c}$} decrease with t, but saturates to a finite value for at least two decades ( SI. Figs S5 for $f=150$ and S6 for $f=300$), implying that the deviations of  dynamical behaviors from normal liquid persist till very long time. The fitted curves are shown for representative cases in insets Fig. 3. Inset, Fig.3(a) shows linear dependence of the Gaussian width parameter $\sigma_{\bot}^{2} (t)$ with time for $f=50$ as in normal liquids. Inset, Fig. 3(b) shows the parameter for spatially exponential tail with parameter $\lambda _{\bot } (t)$ $\sim$ $t^{0.5} $ along with $\sigma_{\bot}^{2}(t)$ $\sim$ $t$ (data not shown) for $f=150$ characterizing non-Fickian diffusion \cite{ng}. The slopes of width parameters of the double Gaussian, $\sigma _{\bot (1)}^{2} (t)$  and $\sigma _{\bot (2)}^{2} (t)$ for $f=300$ show linear dependence on $t$ in Inset, Fig. 3(c).

The in-plane distinct-vHfs, $G_{D}^{(++)} (r_{\bot},t)$ and $G_{D}^{(+-)} (r_{\bot},t)$, are shown in Fig.4(a). The wave-vector ($q_{\bot}$) dependent distinct-vHfs, $G_{D}^{(++)} (q_{\bot},t)$  corresponding to $G_{D}^{(++)} (r_{\bot},t)$ show a peak and $G_{D}^{(+-)} (q_{\bot},t)$ corresponding to $G_{D}^{(+-)} (r_{\bot},t)$ show a dip at wave-vector $q_{\bot}=q_{0}$ [Inset Fig.4(a)]. The structural relaxations are indicated by the decays in the first peak or the dip. We quantify the decay by  $C_{0}^{(++)}(t)[= \frac{G_{D}^{(++)} (q_{0},t)-1}{G_{D}^{(++)} (q_{0},t=0)-1}]$ and $C_{0}^{(+-)}(t)[=\frac{1-G_{D}^{(+-)} (q_{0},t)}{1-G_{D}^{(+-)} (q_{0},t=0)}]$ shown in Fig. 4(b) in semi-logarithmic plots. The relaxation of $+ve$ particles in the neighbourhood of other $-ve$ particles ($C_{0}^{(+-)} (t)$) is slower than that in the vicinity of other $+ve$ particles ($C_{0}^{(++)} (t)$). We observe that the decay is exponential in $t$ in general characterizing diffusive relaxation \cite{hm}. However, $-lnC_{0}^{(+-)}(t) \sim t^{0.75}$ for $f=150$, indicating stretched exponential relaxation. The slow relaxation can be understood within the Vineyard approximation \cite{hm} where $C_{0}(t)\approx G_{S} (q_{0},t)$, the Fourier transform of $G_{S} (\Delta r_{\bot},t)$. Using the asymptotic form where $r_{c}$ is almost independent of $t$, $ln G_{S} (q_{0},t) \sim -t^{0.75}$, if $G_{S} (\Delta  r_{\bot},t)$ has exponential tail (See SI Note.1). 

\begin{figure}[h]
\includegraphics[angle=0,scale=0.07]{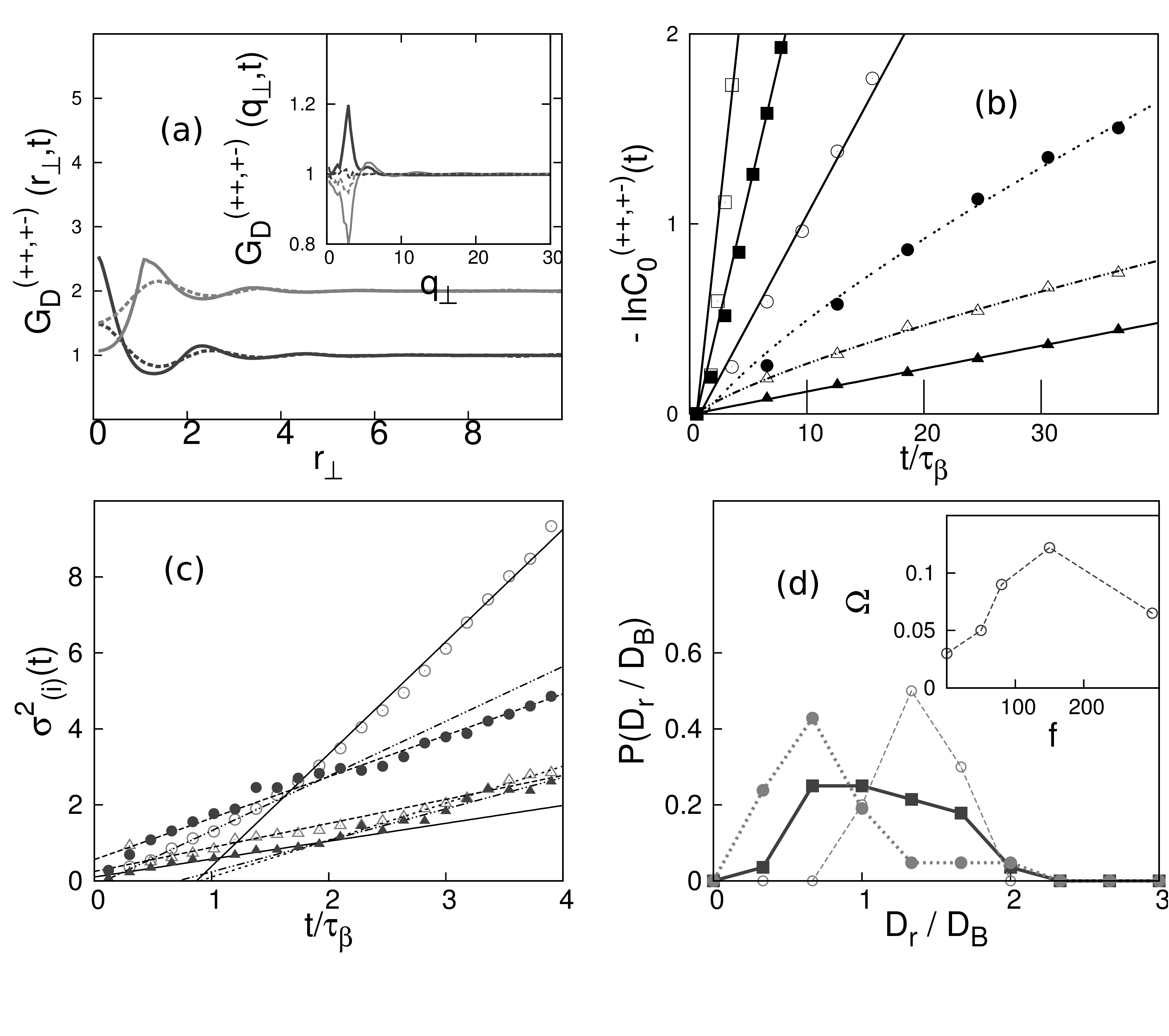}
\caption{ (a) $G_{D}^{(++)} (r_{\bot},t)$ (black) and $G_{D}^{(+-)} (r_{\bot},t)$ (gray)(with vertical offset 1.0) as a function of $r_{\bot}$ for $t=0$ (bold line) and $t=10$ (dotted line) for $f=150$ (Main Panel). Dependence of $G_{D}^{(++)} (q_{\bot},t)$ (black) and $G_{D}^{(+-)} (q_{\bot},t)$ (grey)for $f=150$ for $t=0$ (bold line) and $t=10$ (dotted line) (Inset). (b) Dependence of $-ln C_{0}^{(++)}(t)$ (open symbols) and 
 $-ln C_{0}^{(+-)}(t)$ (filled symbols) for $f=50$ (squares), $f=150$ (circles) and $f=300$ (triangles). Lines show the fitted curves.   (c) $\sigma ^{2}_{(i)}(t)$ vs $t$ plots for two randomly picked particle for $f=150$ (open symbols) and $f=300$ (filled symbols). (d)$P(\frac{D_{r_{} } }{D_{B} } )$ vs $\frac{D_{r_{} } }{D_{B} } $ plots for $f=$ 50 (Dotted line with Circles), 150 (Bold Line with Filled Squares) and 300 (Dotted line with filled circles) Inset, $\Omega$ as a function of $f$.}
\end{figure}

{\it Particle Resolved Picture:} In order to gain microscopic understanding of the dynamic behavior we probe the single particle probability distribution of displacements $P^{(+)(i)}$ $(\Delta r_{\bot},t)$ for tracked $+ve$ 40 particles randomly without any bias to structural regions to represent the statistical behavior of the system \cite{stat}. The second moment, $\sigma _{(i)}^{2} \left(t\right)=\int {\Delta r_{\bot }}^{2} P^{(+)(i)}$ $(\Delta r_{\bot }^{} ,t)d^{2} (\Delta r _{\bot }) $, shown in Fig. 4(c) for $f=150$ and $300$ respectively, has different slopes for different particles in the given time window, the slope being the self diffusion coefficient, $D_{r}$. Fig.4(d) shows the distribution $P(D_{r} /D_{B} )$ for the tagged particles, $D_{B}$ the bulk diffusion coefficient. We observe a sharp peak in $P(D_{r} /D_{B} )$ for $f=50$. The distribution is much broader for $f=150$ but gets sharp again at $f=300$. We consider the width of $P(D_{r} /D_{B} )$ around the peak, $\Omega$, as the measure of heterogeneity in diffusion which exhibits a maximum around $f=150$ (Inset. Fig. 4(d)). 

The exponential tail is shown in self-vHf when $P(D_{r} /D_{B} )$ is broad, in agreement to the phenomenological propositions of Ref. \cite{ng}. The slowing down in dynamic responses is observed in super-cooled systems as well\cite{smk,slow}. However, the slow dynamics in such systems is due to caging of the particles by the neighbours\cite{cage} . In contrast, the individual particle motions are always diffusive in our system. The heterogeneity in diffusion implies that in transport processes. While the mean diffusion is not very meaningful for $f=150$, the mean diffusion for $f=300$ is less than that for $f=50$, qualitatively similar to the lock-in mechanism in Ref. \cite{nat7a}. The distribution of diffusivities depends non-monotonically on the structural correlation length scale which grows with $f$. 

{\it Generalization:} This non-monotonic dependence can be understood in a general context. Let us consider two domains consisting of particles, denoted as $A$ and $B$, having a relative drift, $v_{rel}$ in $z$-direction. Let $V^{(AA)}(r)$ be the mutual interaction of the particles in a domain $A$ and $V^{(BB)}(r)$
that in a domain $B$, while $V^{(AB)}(r)$ that between particles of different domains. The local density in the domain changes due to single particle motion. Let us focus one of the domains, say domain $A$. The single particle motions are: (1) Diffusive and drive currents of particles $A$ in the given domain; (2) Movement of particles $A$ to domain $B$. The second type of movement across different domains costs interaction energy due to changes in the neighboring particles, $\delta V \approx [V^{(AA)} (r=\rho _{0}^{-1/3} )-V^{(AB)} (r=\rho _{0}^{-1/3} )]\rho_{0}$ at mean density$\rho _{0} $ where $\rho _{0}^{-1/3}=$ the mean separation between the particles. The probability of inter-domain movements,  $p\sim exp(-\delta V/k_{B}T)$. These particles will experience a relative force of magnitude, $f_{rel}(=\Gamma v_{rel})$ corresponding to the relative drive, having energy $\delta V_{ext} \sim pf_{rel}\rho^{-1/3}$. This contributes a current $-[v_{rel} \exp \left(-\delta V_{ext}/k_{B} T\right)] \delta \rho ^{(A)} \left(\vec{r}_{\bot } ,z,t\right)$, with $\delta \rho ^{(A)} \left(\vec{r}_{\bot } ,z,t\right)$= $\rho ^{(A)} \left(\vec{r}_{\bot } ,z,t\right)$-$\rho _{0} $ where $\rho ^{(A)} \left(\vec{r}_{\bot } ,z,t\right)$ is local density in a domain. The negative sign accounts for the loss of particles from domain $A$. The equation of density profile in the steady state,
\begin{eqnarray}
\nonumber
[D\nabla _{\bot }^{2} +v_{rel} (1-\exp \left(-\delta V_{ext}/k_{B} T\right))\partial _{z} ]\delta \rho ^{(A)} (\vec{r}_{\bot } ,z,t)\\
 =N(\vec{r}_{\bot } ,z,t) ~~~~~~ \label{de}
\end{eqnarray}
where $N\left(\vec{r}_{\bot } ,z,t\right)$ is noise with Gaussian statistics. The diffusive current in $z$ has been ignored in comparison to the drive current. The second term represents competition between drive and particle interaction. The density correlations in the transverse plane in Eq. \eqref{de} decays as $\exp (-\lambda 'r_{\bot } /D)$, where $\lambda '=v_{rel} [\exp (-\delta V_{ext}/k_{B} T)-1]$ (SI Note SN. 2).

The above mentioned scenario assumes movement of a particle with diffusion $D$. Let us now consider a situation of pair of particles. However, $D$ is not fixed but having a distribution as suggested by our simulations. We take the diffusion coefficients $D$ and $D+\delta D$ where $\delta D$ is a random variable over the pairs. The relative diffusion of the particles $\tilde{D}= 2D+\delta D$. If the particles are structurally uncorrelated, the separation between a pair of particles fluctuates with time with the probability $\sim$ $ \exp (-r_{\bot }^{2} /4\tilde{D}t)$. Due to structural correlation, the probability is given by $\sim \exp (-\lambda 'r_{\bot } / \tilde{D})$ $ \exp (-r_{\bot }^{2} /4\tilde{D}t)$. Integrating over $r_{\bot}$ and $t$  results in moments of probability distribution function of $\delta D$, $ln P(\delta D)\sim -(\delta D/ \omega)^{2}$ where $ {\omega\sim v_{d} ^{3} \exp \left(-\delta V_{ext}/k_{B} T\right)}$ (SI Note SN.3).

For our simulated system, $A$ and $B$ stands for positively and negatively charged particles respectively. Using $V^{(AA)}=V^{(++)}\approx \frac{V_{0}} {(1+\frac{\kappa }{2})^{2}} [\frac{\exp(-\kappa(r -1))}{(r)}]$, $V^{(AB)}=V^{(+-)}\approx \frac{-V_{0}} {(1+\frac{\kappa }{2})^{2}} [\frac{\exp(-\kappa(r -1))}{(r)}]$ and $v_{rel} \approx v_{d}= f/ \Gamma$ with the parameters taken from simulation model, we obtain the maximum in $\omega$ at $f \approx 130$. From simulation, we observe maximum in $\Omega$ at $f\approx 150$ close to $f_{C}$. Thus the model can qualitatively captures the anomalous issues we found in the BD simulation. 

Our model is directly applicable to phase separating liquids in presence of drive \cite{lab,colr}. Also, there exist similar condensed matter systems, where anomalies in dynamical responses have been described phenomenologically to heterogeneity in diffusion \cite{ng,smk,strrel1,ng1,ng2}. The generality of our analysis points to exploring the possibility of relating such heterogeneity to non-equilibrium effects where the particle interactions compete with local drive \cite{strrel,colr,lab}. The local drive need not be necessarily external, but may arise out of unbalanced thermodynamic forces due to non-uniform distribution of pressure, temperature and chemical potential in the system.

To conclude we show, using BD simulations, exponential tail in self-vHf and stretched exponential structural relaxation due to heterogeneity of diffusion in non-equilibrium steady states. Our theoretical analysis show that the anomalous dynamic behaviour is due to competition between the particle interaction and drive in the system. These results can be verified by experiments on colloid. The heterogeneity in diffusion reflects heterogeneous transport which is tunable by external drive. This may be harnessed in  technological applications \cite{colr,lab}. Moreover, our analysis is general enough to throw light to microscopic origin of heterogeneous diffusion. 

We thank Prof. C. Dasgupta and S. Karmakar for insightful discussions.

%%%%%%%%%%%%%%%%%%%%%%%%%%%%%%%%%%%%%%%%%

\begin{thebibliography}{}

%\bibitem{mcdermott}
 % Y.-F.~Chen {\it et al.}, Phys.\ Rev.\ Lett.\  {\bf 107}, 217401 (2011).
\bibitem{hm}
J. P. Hansen, and I. R. McDonald, Theory of Simple Liquids, ( Academic Press, London, 1986).

\bibitem{cl}
P. M. Chaikin, T. C. Lubensky, Principles of Condensed Matter Physics, (Cambridge University Press, 2000)

\bibitem{soft1}
R. A. L. Jones,  Soft Condensed Matter, Oxford Master Series in Physics(Oxford University Press, Oxford, 2002)

\bibitem{soft2}
D. David Andelman, and G. Reiter, (Ed) Series in Soft Condensed Matter, Vol-1-6, (World Scientific, Singapore, 2012).

\bibitem{hoh}
M. C. Cross, P. Hohenberg, ,  Rev.\ Mod.\ Phys.\ ,{\bf 65}, 851 (1993)


\bibitem{nat3}
H. L{\" o}wen, Phys.\ Rep.\ , {\bf 237}, 249 (1994)


\bibitem{nat3a}
H. L{\" o}wen,  J.\ Phys.\: Condens.\ Matter\ , {\bf 13}, R415-R432 (2001)


\bibitem{nat4}
A. V. Blaaderen,  {\it et al.},  Eur.\ Phys.\ J.\ Special\ Topics, {\bf 222,} 2895-2909 (2013)

\bibitem{nat4a}
H. L{\"o}wen,  Eur.\ Phys.\ J.\ Special\ Topics, {\bf 222}, 2727-2737 (2013)

\bibitem{nat5}
J. Chakrabarti, J. Dzubiella, H. L{\" o}wen,  Europhys.\ Lett.\ , {\bf 61}, 415 (2003)

\bibitem{nat5a}
J. Chakrabarti, J. Dzubiella, and H. L{\" o}wen, {\it Phys.\ Rev.\ E}, {\bf 70}, 012401 (2004)

\bibitem{nat6}
M. E. Leunissen,  {\it et al.}, Nature, {\bf 437}, 235-240 (2005)

\bibitem{nat6a}
M. Rex,  H. L{\" o}wen,  Phys.\ Rev.\ E,  {\bf 75}, 051402 (2007) 

\bibitem{nat6b}
K. R. S{\"u}tterlin, {\it et al.},  Phys.\ Rev.\ Lett., {\bf 102}, 085003 (2009)

\bibitem{nat7a}
T. Vissers, {\it et al.}, Soft Matter, {\bf 7}, 2352 (2011)

\bibitem{nat7b}
T. Glanz, and H. L{\"o}wen, J.\ Phys.:\ Condens.\ Matter\ , {\bf 24}, 464114 (2012)

\bibitem{vhv}
L. Van Hove, Phys.\ Rev.\ {\bf 95}, 249 (1954)

\bibitem{smk}
S. Sengupta, and S. Karmakar, J.\ Chem.\ Phys., {\bf 140}, 224505 (2014)

\bibitem{slow}
S. Karmakar, C. Dasgupta and S. Sastry, Annu.\ Rev.\ Condens.\ Matter.\ Phys {\bf 5}, 255 (2014)


\bibitem{cage}
K. N. Pham {\it et al.}, Science, {\bf 296}, (2002)

\bibitem{ng}
B. Wang,  J. Kuo,  S. C. Bae,  S. Granick, Nat.\ Mat., {\bf 11}, 481 (2012).


\bibitem{erm}
D. L. Ermak,  J.\ Chem.\ Phys. , {\bf 62}, 4189 (1975)


\bibitem{stat}
J. A. Greenwood, and M. M. Sandomire, J.\ Amer.\ Statist.\ Assoc\, {\bf 250}, 257-260 (1950)

\bibitem{lab}
N. Cevheri and M. Yoda, Lab Chip {\bf 14}, 1391 (2014)

\bibitem{colr}
I. S. Aranson, Phys.\ -Usp. {\bf 56}, 79(2013)



\bibitem{strrel1}
E. E. Ferrero, K. Martens and J.-L. Barrat, Phys.\ Rev.\ Lett.\ , {\bf 113}, 248301 (2014);


\bibitem{ng1}
Y. Guan, B. Wang, and S. Granick,  {\it ACS Nano}, {\bf 8}, 3331-3336 (2014)

\bibitem{ng2}
G. Kwon, B. J. Sung, and A. Yethiraj, J. Phys. Chem. B, {\bf 118}, 8128-8134 (2014)


\bibitem{strrel}
K. Paeng {\it et al.}, Proc.\ Natl.\ Acad.\ Sci.\ USA\ , {bf 112} 4952(2015);  R. Torre, P. Bartolini and R. Righini, Nature, {\bf 428}, 296 (2004); Pablo G. Debenedetti and Frank H. Stillinger, Nature, {\bf 410} 259 (2001)




\end{thebibliography}
\end{document}